# Passive Exposure to Mobile Phones:
# Enhancement of Intensity by Reflection

Tsuyoshi HONDOU*, Takenori UEDA[1], Yasuhiro SAKATA[2], Nobuto TANIGAWA[2],
Tetsu SUZUKI[3], Taizo KOBAYASHI[2] and Kensuke IKEDA[2]

*Department of Physics, Tohoku University, Sendai 980-8578*
*[1]Japan Offspring Fund, 2-5-2 Kojimachi, Chiyoda-ku, Tokyo 102-0083*
*[2]Department of Physics, Ritsumeikan University, Kusatsu 525-8577*
*[3]Department of Information and Communication Engineering,*
*Sendai National College of Technology, Sendai 989-3128*

In a recent Letter [J. Phys. Soc. Jpn. **71**, 432 (2002)], we reported a preliminary calculation and concluded that public exposure to mobile phones can be enhanced by microwave reflection in public spaces. In this paper, we confirm the significance of microwave reflection reported in our previous Letter by experimental and numerical studies. Furthermore, we show that "hot spots" often emerge in reflective areas, where the local exposure level is much higher than average. Such places include elevators, and we discuss other possible environments including trains, buses, cars, and airplanes. Our results indicate the risk of "passive exposure" to microwaves.

### §1 Introduction

Significant public concerns have arisen over exposure to microwaves. These concerns relate to interference with electro-medical devices and the effect on human health. Exposure to microwaves has been shown to disturb the regular activity of our biological cells, even at exposure levels much lower than those known to induce thermal stress. It is naively assumed that in public areas the exposure level decreases rapidly with distance from the radiation source (e.g., a mobile phone)[1-10]. This assumption has been widely applied to several risk assessments[11-13].

From the viewpoint of mathematical physics, the laws of electromagnetic waves are described by Maxwell's equations, a set of partial differential equations. It is well known in mathematics that a partial differential equation cannot be solved without specifying its boundary condition. In the application we present here, microwaves are reflected by metallic walls at an efficiency of more than 99%. Hence, we recently pointed out in preliminary calculations[14] the importance of considering all boundary condition in passive exposure to microwaves, since many reflective boundaries are found in our daily life, for example, in elevators and trains. However, specialists in this area have still not acknowledged the central issue raised.

A typical objection to concerns about reflected microwaves was presented in the magazine *New Scientist*, in which a specialist conceded that microwaves would bounce around the inside of carriages and boost field levels. However, he claimed, "the increase should be minimal, because power drops off a short distance away from each phone"[15]. Another expert also stated, "RF emissions are not "trapped" in a train car. They disperse, are absorbed and dissipate in intensity."[16]

It seems that specialists with this viewpoint are not familiar with the basics of partial differential equations, and that in their public exposure assessments[11-13] they are assuming a boundary condition under which there is no reflection. Thus, we "quantitatively" studied the effect of the boundary

---

* E-mail address: hondou@cmpt.phys.tohoku.ac.jp
  Web address: http://www.cmpt.phys.tohoku.ac.jp/~hondou/

condition to identify whether it is necessary to account for the effect of microwave reflection in evaluating the potential health hazard of microwave exposure in daily situations. The aim of this paper is to determine whether Case 1 or Case 2 is correct:

**Case 1)** The effect of reflection is negligibly small. The effect of reflection may not be hazardous.

**Case 2)** The effect of reflection is NOT negligible. We have to seriously consider the effect of reflection, leading to "passive exposure" in public areas.

In this paper, we report experimental and numerical results demonstrating that Case 2 holds. Because of microwave reflection, the exposure level can be orders of magnitude higher than that conventionally assumed by specialists. In §2, we describe the experimental setup and the results of microwave distributions in a container and an elevator. In §3, we describe numerical simulations that confirm the experimental results. In §4, we summarize our results and discuss the significance of the interdisciplinary research from diverse viewpoints of the problem.

## §2 Experiments

### 2.1 Experimental methods in container

For our experiment, we used a 20-foot metallic reefer (refrigerator) container. The inner wall surfaces of the container were made of stainless steel while the ceiling and floor were aluminum. All the metal surfaces reflect microwaves at an efficiency of greater than 99.99%. The internal dimensions of the container were 5.5 m (length) × 2.3 m (width) × 2.2 m (height), which are represented as a three-dimensional (3D) coordinate system as shown in Fig. 1 (X: [0, 5.5], Y: [0, 2.3], Z: [0, 2.2]).

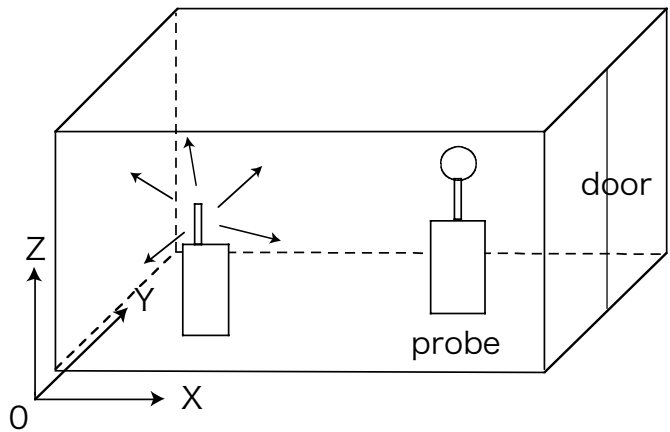

Figure 1 Installation of radiation source (FM transmitter) and probe.

As a microwave source, we installed a 1.2 GHz frequency modulated (FM) radio transmitter (TH-59, Kenwood, Tokyo) in the container at X=0.69, Y=0.66, Z=1.12. The frequency of the transmitter was within the frequency range of mobile phones. In the experiments, the transmitted power was about 0.15 W. The peak radiation power of a mobile phone may reach around 2 W depending on the model and conditions. Because the radiation power of actual mobile phones fluctuates with time depending on their operating condition, we used an amateur radio transmitter to fix radiation power. The transmitter was supported by a cardboard box, whose absorption of microwaves was known to be negligible. A rod antenna included in the transmitter set from the manufacturer was used.

The intensity distribution of the microwaves was measured by a hand-held isotropic electric-field probe put in front of an observer. The existence of an observer perturbs the microwave distribution, which generally decreases the intensity by absorption, but may increase the local intensity by reflection[17]. A pilot study between two cases (with and without an observer) showed that the effect of an observer on the overall microwave distribution is negligible compared with the order of enhancement of intensity by reflection. An electric field probe (Type 18.0) and electromagnetic radiation meter (EMR-21) from Narda Safety Test Solutions (Pfullingen, Germany) were used for all experiments. The equipment was calibrated at Narda on May 7, 2003 (probe) and May 9, 2003 (meter). A calibration factor of 0.92 (at 1.2 GHz and 1.4 GHz) was used in the experiments.

### 2.2 Experimental result in container

A typical result of the power density (equivalent Poynting's vector[18]) distribution is shown in Fig. 2. We also plot the intensity distribution as typically assumed by specialists of radio engineering, where the intensity is approximated as an inverse square law. The experimental values of intensity are consistently higher than the predicted values. Intensity does not even decrease with distance from the source. The difference in intensity between the container situation and the free boundary condition increases with distance. The higher intensities measured under a reflective boundary condition are simply a consequence of the law of energy conservation.

We furthermore confirm the existence of microwave "hot spots", in which the microwaves are "localized". The intensity measured at one hot spot 4.6 m from the transmitter is the same as that at 0.1 m from the transmitter in the case without reflection (free boundary condition). Namely, the intensity at the hot spot is increased by approximately 2000 times by reflection. The spatial fluctuation of the intensity is attributable to the wave nature of electromagnetism, in which the phases of electromagnetic waves coming from an infinite number of paths in a 3D space interfere with each other. The interference leads to an intensity distribution that is highly sensitive to an individual's position. We performed experiments in different situations: 1) without people in the container, 2) with the door open without people, and 3) with people in the container. Indeed, the average intensity depends on the condition. However, the following characteristics of the intensity do not change: 1) a higher-than-predicted intensity due to reflection, 2) spatial fluctuation of the intensity, and 3) the existence of hot spots.

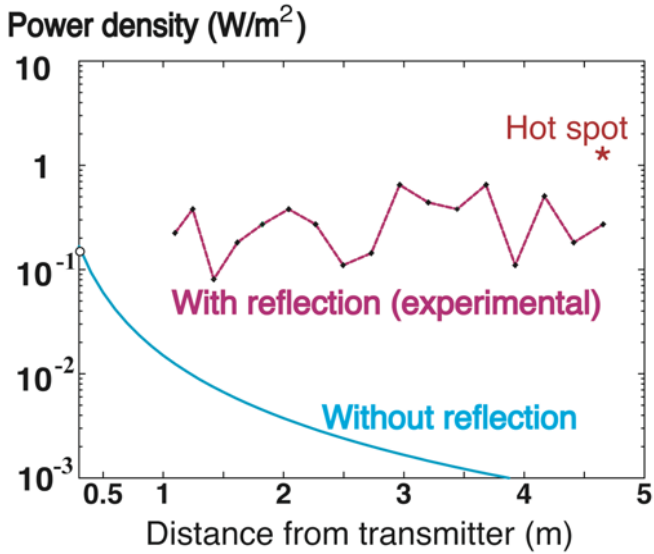

Figure 2 Distribution of microwave intensity in container, for which the data were obtained along the X-axis: X: [0.75(m), 5.25(m)] with fixed Y and Z coordinates: Y=1.60 (m) and Z=1.00 (m). The sequential line graph titled "With reflection" shows the measured intensity when following a straight path along the X-direction in the container. We plot the power density of the wave zone only, in which the distance from the radiation source is sufficiently far relative to the microwave wavelength. The "Without reflection" curve shows the expected intensity in the case without reflection determined by an inverse square law from the measured intensity at 0.30 (m) in the free boundary condition. A hot spot caused by the localization of electromagnetic waves occurs at X=5.30, Y=0.07, Z=1.00.

### 2.3 Experiment in elevator

The results shown in the container is closely connected to our everyday life. Elevators comprised of metal are a typical and concrete example, in which almost all the boundary reflects electromagnetic waves. We consider the container an effective model of an actual elevator, because the difference in reflection rates between our container surface and those of metallic elevator surfaces (steel in most cases) is negligibly small. However, we performed further experiments to indicate the universality of our result. In experiments at Tohoku University, Japan, we installed the same FM transmitter in an elevator with a capacity of 17 people (1.5 m × 1.8 m × 2.3 m). With 2 people in the elevator, we observed a hot spot of 1.1 (W/m²) at a distance 2.6 m from the transmitter *even when the door was open*. This intensity is realized in the immediate vicinity of the transmitter, i.e., at 0.1 m, in the free boundary condition.

### § 3 Numerical Simulations

To confirm our experimental findings of the greater-than-predicted intensity due to reflection, as well as the hot spots, we performed two numerical simulations using High-Frequency Structure Simulator (HFSS) version 9 from Ansoft Corporation (Pittsburgh, U.S.A.) and Finite-Difference Time-Domain (FDTD) methods, which are the most standard and reliable methods for the analysis of high-frequency electromagnetic fields (for detail, see Appendix).

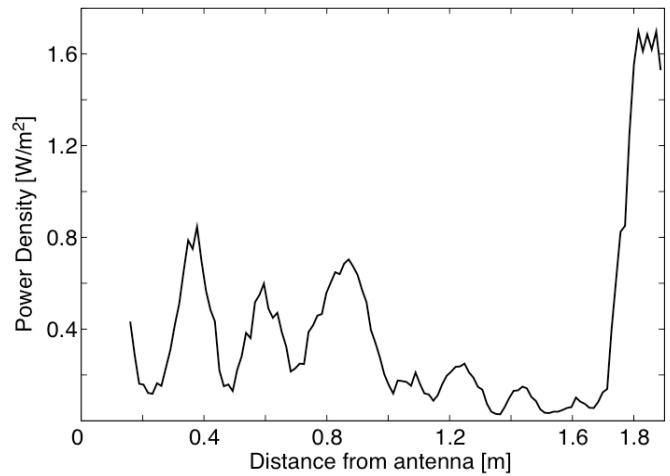

Figure 3 Intensity distribution in the elevator model, obtained using HFSS, in which the data was obtained along the line between (0.50, 0.50, 0.11) and (1.48, -0.80, 1.07). The dimensions are the same as those used in the experiment, except for the reduced height of 1.3 m from 2.3 m due to limited CPU memory, i.e., (1.5 m × 1.8 m × 1.3 m). The door of the elevator is kept fully open, and the width of the aperture is 0.9 m. The intensity of the hot spot at 1.8 m is increased by approximately 1000 times by reflection.

Figure 3 shows a typical result of the intensity distribution in an elevator model, obtained using HFSS, in which the door of the elevator was *fully ope*n, as in the experiment. A snapshot of the spatial distribution of the intensity[19] is shown in Fig. 4, in which the element of the antenna is shown by a black bar and the intensity is indicated by a color scale. The intensity does not monotonically decrease from the transmitter, which is in clear contrast to the case without reflection[20] (Fig. 5), in which all the parameters are the same as those shown in Fig. 4 with the exception of the boundary condition. The intensity at the hot spot (X, Y, Z) = (1.46, -0.78, 1.05) around 1.8 m from the transmitter in the reflective boundary condition is approximately 1000 times higher than that at the same position in the free boundary condition[21]. The result of the simulation is thus consistent with our experiments, although the values differ owing to the different conditions imposed by computational limits.

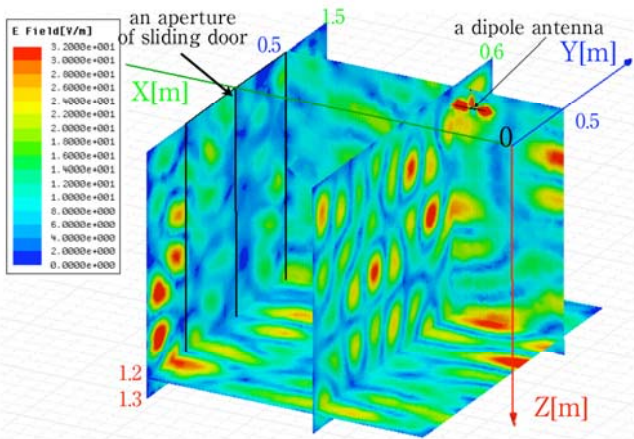

Figure 4 Snapshot of spatial distribution of intensity in elevator model, obtained using HFSS. Parameters are the same as those shown in Fig. 3. The door of the elevator, whose area is specified by X=1.5, Y:[-0.45, 0.45] and Z:[0, 1.3], is kept fully open. For details, see Appendix and ref. 19.

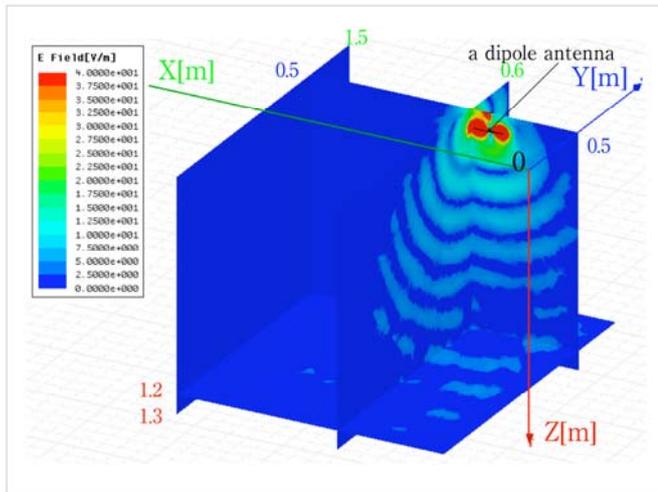

Figure 5 Snapshot of spatial distribution of intensity in the case without reflection. Parameters are the same as those for the elevator model, except for the boundary condition. For details, see Appendix and ref. 20.

The experiment in the container was numerically simulated by a two-dimensional (2D) FDTD[22] (Fig. 6), which is also consistent with the experiments and numerical results using HFSS. The 2D simulation does not correspond directly to a real experiment in 3D space because of the difference in dimensionality. However, the result of the experiment is also reproduced: a greater than predicted intensity due to reflection, as well as the existence of hot spots. For example, we observe a hot spot at (X, Y) = (432, 72) cm. In comparison with the control simulation using the free boundary condition, we find that the power density at the hot spot is increased by approximately a thousand times by reflection.

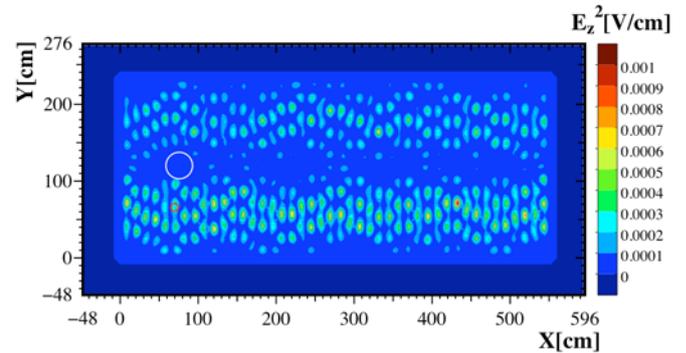

Figure 6 Snapshot of spatial distribution of intensity in 2D container model simulated by FDTD method. We set a radiation source and a human model as (X,Y)=(70,66) and (76,120), respectively. For details, see Appendix and ref. 22.

## § 4 Summary and Discussion

These experimental and numerical results clearly reject Case 1 and support Case 2, that is, microwave reflections must be considered in risk assessments of electromagnetic exposure in daily life. We have already noted that in elevators, the passive exposure level is much higher than previously thought. Other places in which passive exposure is likely to be enhanced include buses, trains, taxis, and airplanes. With enhanced exposure levels caused by reflection, we may be passively exposed beyond the levels reported for electro-medical interference and health risks. For example, one report showed that mobile phone radiation interfered with heart pacemakers from 30 cm away[13], even without reflection. The present result also confirms evidence that some hearing aids often suffer from heavy noise caused by microwave radiation in public transport.

In relation to biological and health aspects, there are several non-thermal effects of microwaves as found in a review by Hyland[1,2]. For example, Salford et al.[3] reported that the blood-brain barrier in rats opens in response to a mobile phone brought to within 1.8 m, even without reflection. Such non-thermal biological effects include DNA strand breaks[4,5], gene expression changes[5,6], allergies[7] and electromagnetic hypersensitivity[8]. If these consequently turn out to be serious for our health, we may be subject to adverse health effects even without owning a mobile phone. Recent epidemiological studies indicate an increased tumor risk after a latency period on the order of ten years[9,10]. Hands-free phone kits may be useful for the users themselves, but may still enhance passive exposure to others who have no control over the exposure.

In the present study, we restricted experiments and numerical simulations to one transmitter for simplicity. In public situations, several mobile phones can be in operation simultaneously, which may further increase the passive exposure level. Because the *peak exposure level* is crucial in considering electro-medical interference, interference to airplanes, and biological effects on human beings, we also need to consider the possible peak exposure level, or hot spots, for the worst-case estimation. Thus, risk assessments based only on average exposure[23-25] are inappropriate[26].

One can avoid exposure to microwaves by not using mobile phones. However, people cannot avoid passive exposure from others, which is a parallel situation to passive smoking. Exposed people include children, babies and, in particular, fetuses, who are likely to be the most sensitive to environmental stresses. Discussion and further research of passive exposure risks under various conditions are seriously encouraged.

## Acknowledgements

The authors thank Yusaku Kimura for assistance in the container experiments at Kawasaki port, Masaki Sano for making us aware of several theoretical aspects of these experiments, and Evan Blackie for critical reading of the manuscript. The authors thank the Yukawa Institute for Theoretical Physics at Kyoto University. Discussions during the YITP workshops YITP-W-03-01, 04-01 and 05-03 on "Biological effects of electromagnetic field" were useful in completing this work. This work was supported in part by a Japanese Grant-in-Aid for Science Research Fund from the Ministry of Education, Culture, Sports, Science and Technology (Grant No. 17654082).

## Appendix: Numerical Method

### 1) Numerical Simulation using HFSS

The elevator simulation was performed using HFSS version 9 from Ansoft Corporation (Pittsburgh, U.S.A.). HFSS utilizes a 3D finite element method (FEM) to solve Maxwell's equations. HFSS solves Maxwell's equations for the stationary state. In this simulation, we selected a "Low-order solution" in HFSS to calculate the largest system volume possible.

Because of limited CPU capacity, a 3D simulation of the actual container was not practical. Instead, to reduce the demand on memory, we performed a numerical simulation for an elevator model with the height reduced from 2.3 m to 1.3 m (i.e., depth, width and height of 1.5 m, 1.8 m and 1.3 m, respectively). The door of the elevator was kept *fully open in all simulations*.

As a radiation source, we installed a 1/2-wavelength dipole antenna. The radiation frequency was 0.9 GHz. This difference in frequency between the experiment (1.2 GHz) and the simulation was also necessary due to memory limitations. The input power was 1 W. All the surfaces were set as stainless steel except the door aperture of the elevator, which was set to be fully open. The boundary condition of the opened door was set to "radiation" for the HFSS simulation.

In the movies, X:[0, 1.5], Y:[-0.9, 0.9] and Z:[0, 1.3] coordinates correspond to the depth, width and height, respectively. The area specified by X=1.5, Y:[-0.45, 0.45] and Z:[0, 1.3] is the open door (the door frame is indicated by the thick lines in Figs. 4 and 5), through which microwaves dissipate out of the elevator. The antenna is set at (X, Y, Z) = (0.50, 0.50, 0.11).

We also have the "project files" for the HFSS simulations. Any user of HFSS (version 9 and later) can verify the present result using the project files (in refs. 27 and 28).

### 2) Numerical simulation by 2D-FDTD

FDTD is a widely used method for solving Maxwell's equations in the time domain based on the calculus of finite differences. FDTD solves the equations not only for stationary states but also for transient processes; however, we only discuss the stationary state here. We performed 2D-FDTD simulation because this method enables us to simulate a large area such as a container. We used the same frequency as that used in the experiment (1.2 GHz).

In Fig. 6, the red circle represents a radiation source. The white circle represents an observer model of radius 10 cm that interacts with the radiation, having the same dielectric constant and conductivity as those of human skin[29].